# Correlation between radiation processes in silicon and long-time degradation of detectors for high energy physics experiments

Sorina Lazanu,[a,] Ionel Lazanu,[b]

[a] *National Institute for Materials Physics, POBox MG-7, Bucharest-Măgurele, Romania*

[b] *University of Bucharest, Faculty of Physics, POBox MG-11, Bucharest-Măgurele, Romania*

Abstract

In this contribution, the correlation between fundamental interaction processes induced by radiation in silicon and observable effects which limit the use of silicon detectors in high energy physics experiments is investigated in the frame of a phenomenological model which includes: generation of primary defects at irradiation starting from elementary interactions in silicon; kinetics of defects, effects at the p-n junction detector level. The effects due to irradiating particles (pions, protons, neutrons), to their flux, to the anisotropy of the threshold energy in silicon, to the impurity concentrations and resistivity of the starting material are investigated as time, fluence and temperature dependences of detector characteristics. The expected degradation of the electrical parameters of detectors in the complex hadron background fields at LHC & SLHC are predicted.

## 1. Introduction

Semiconductor crystals possess a high sensitivity to radiation. The action of radiation on crystals leads to the formation of lattice defects and to substantial changes in their properties. Intensive investigation of their behaviour under irradiation has been conducted over a period of several decades, but, up to now, not all physical aspects are clarified.

The motivation of this study comes from the fact that silicon detectors will represent an option for the next experiments in high energy physics where the requirement to work long time in high radiation environments will represent a major problem.

In this contribution, the correlation between fundamental interaction processes induced by irradiation in silicon, microscopic aspects of the degradation (production and time evolution of the concentration of defects) and observable effects at the level of the p-n junction detector, which limit the use of silicon detectors: increase of the leakage current, of the effective carrier concentration in the space charge region ($N_{eff}$), decrease of the charge collection efficiency due to the modification of the effective trapping time of charge carriers is investigated. A model for the degradation, developed successively by the authors, is used – see Ref. [1] and references cited therein.

The main hypothesis made in the understanding of fundamental phenomena in silicon is brought by the consideration of the existence of the fourfold coordinated defect [2], as a primary defect in addition to "classical" vacancies and interstitials. This defect introduces a new type of symmetry of the lattice.

The results of the model are compared with available experimental data permitting to estimate the quality of predictions and their sensitivity to different parameters. The effects due to irradiating particles (pions, protons, neutrons), to the damage rate, to the anisotropy of the threshold energy in silicon, to the impurity concentrations, resistivity and growth technology (FZ, DOFZ) of the starting material are considered as time, fluence and temperature dependencies of detector characteristics.

The model could explain in a natural manner the modifications of electrical parameters of devices especially after hadron irradiation.
The time evolution of electrical characteristics of silicon detectors due to the hadron background expected at the LHC and SLHC is predicted.

## 2. Modelling semiconductor degradation due to irradiation

The primary mechanism of defect formation during irradiation in semiconductors is the collision of the incoming particle with the atoms of the crystal, which leads to the departure of the atom to a rather large distance from its original site, i.e., to the formation of separated Frenkel pairs. The process continues as a cascade, up to the threshold energy limit of the energy of recoils. In silicon, the components of the pair - the vacancy *V* and the self interstitial *I* - are mobile over a broad temperature range, including room temperature, and during diffusion they are either trapped by other primary defects and impurities with the formation of secondary defects, or migrate to sinks. The main characteristics of the kinetics of defects are summarised in Ref. [3].

A special mention must be made for the $Si_{FFCD}$ defect. Predicted by Goedecker and co-workers [2], some characteristics were determined by the authors in Ref. [3]. This defect must be more stable than the vacancy for a large temperature interval [4]. The $Si_{FFCD}$ has not yet been detected experimentally, but this fact is not unexpected because the search for defects is always guided by theoretical predictions. Moreira et. al. [5] found that the FFCD exists and is also stable in Ge, and their prediction is in agreement with experimental data [6]. In principle, this new type of defect could be characteristic for all semiconductors with diamond structure.

The knowledge of the kinetics of defects in silicon during and after irradiation permits the determination of the effects produced at the p-n junction detector level: modification of the leakage current, effective carrier concentration in the space charge region ($N_{eff}$) and the effective trapping probabilities of charge carriers.

## 3. Results and discussion

In Figures 1a ÷ c, the fluence dependence of the concentrations of defects, induced by $2 \cdot 10^{14}$ cm$^{-2}$ neutron irradiation, both in FZ and DOFZ silicon, is presented as an example. The calculations reproduce the conditions at 10 days after irradiation, at room temperature. While the concentration of $Si_{FFCD}$ depends linearly on fluence, the concentration of vacancies in FZ Si has a tendency toward saturation. In DOFZ Si, the concentration of vacancies has a linear dependence on fluence, but is several orders of magnitude lower than in FZ Si, due to the fact that oxygen trap vacancies and form the VO centre. The presence of oxygen and the formation of VO and $C_iO_i$ centres, which involve both vacancies and interstitials, conduce to much quicker stabilisation in time of oxygen rich silicon after irradiation. While for FZ materials the concentrations of $V_2$ and $C_iC_s$ have been estimated to be the most important, in DOFZ VO and $C_iO_i$ are the defects with the highest concentrations.

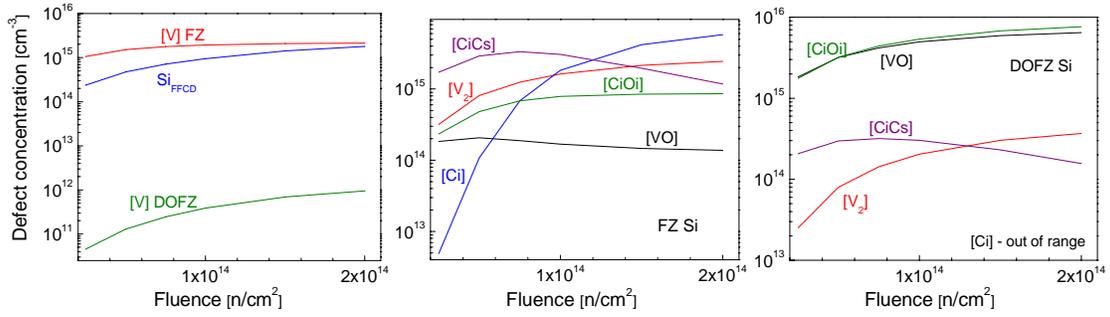

Fig. 1.
Fluence dependence of defect concentrations in FZ and DOFZ Si after irradiation with $2 \cdot 10^{14}$ n/cm$^2$.

The dependencies of the reciprocal of the effective trapping times for electrons (left) and holes (right) on the pion (up) and neutron (down) fluences are represented in Figures 2a ÷ d. Lines correspond to calculations in the frame of the present work, while points are experimental data from Ref [7]. Results similar to those presented for pions have been obtained for protons.

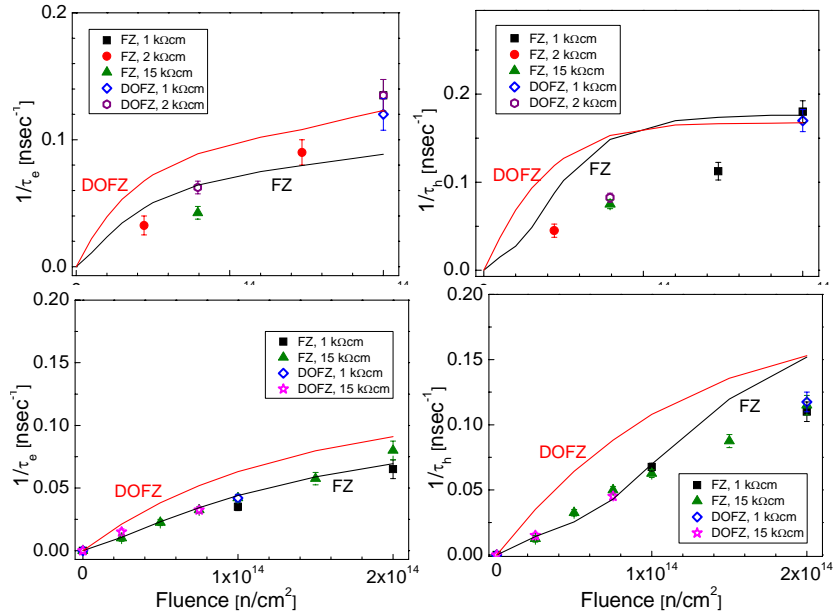

Fig. 2.
Fluence dependence of the effective trapping probability for electrons (left) and holes (right) in FZ (up) and DOFZ (down) Si detectors.

The differences in silicon resistivity in the range 1 to 15 kΩcm have irrelevant influence on the effective trapping times for electrons and holes, but they depend on the type and energy of irradiating particle. The model reproduces very well the experimental effective electron trapping times in FZ material and differences – in a reasonable range exist for DOFZ materials and for hole trapping times. Probably, additional mechanisms in the kinetics of defects must be considered in the model to diminish these discrepancies.

The correlation between fluence, time, temperature, anisotropy of the threshold energy for displacements, initial impurity concentrations in the starting materials and variation of device parameters (leakage current and $N_{eff}$) has also been investigated.

In Figure 3, the time dependence of the degradation constant of the leakage current at 20 and 0 °C is presented; experimental data from Ref. [8] are compared with model calculations: continuous (dashed) curves are associated with calculations with (without) consideration of the contribution of $Si_{FFCD}$.

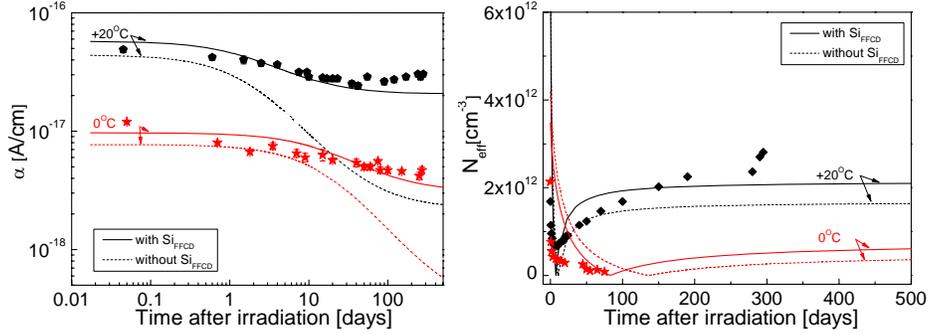

Fig. 3
Time dependence of the: a) degradation constant of the leakage current and b) $N_{eff}$ at 20 and 0 °C.

The fluence dependence of $N_{eff}$ after 24 GeV/c proton irradiation is represented in Figure 4 for silicon detectors with initial resistivity 1 KΩcm: FZ with <100> and <111> orientations, and <111> DOFZ - experimental data are from Ref [9]. Crystal orientation has been introduced into the model by the anisotropy of the threshold energy for displacements. In the figure the dotted curve calculated without the consideration of the $Si_{FFCD}$ defect illustrates its contribution to degradation.

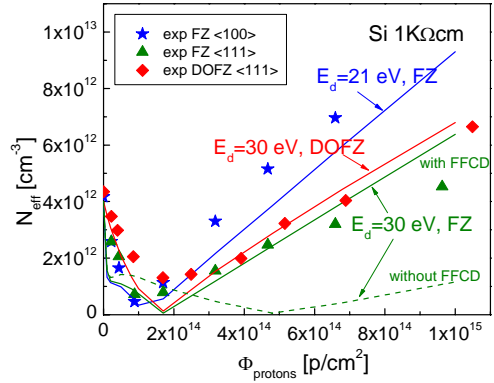

Fig. 4.
$N_{eff}$ versus proton fluence for <100> and <111> FZ and <111> DOFZ Si

All these results put in evidence the essential role of the introduction of the $Si_{FFCD}$ defect in the modelling the degradation processes.

Silicon detectors will be used in experiments at LHC and in its up-grade in the tracking of charged particles. After the long-time operation in these high radiation fields, the interactions in the detector could produce modifications in their parameters, which eventually could attain unacceptable levels.

Predictions of the degradation of silicon detectors in conditions of continuous irradiation, as will be the case of the next experiments in HEP, LHC and SLHC, are presented in Figure 5. Differences in the volume density of the leakage current for FZ and DOFZ detectors exist only at short times (in the first year), and the benefit of oxygenation, for 0 °C operation, is not seen in $N_{eff}$ both in the environment of LHC and of SLHC. In what regards the effective trapping probabilities, these are higher in oxygenated materials. They present a linear dependence for electrons, and a relatively quick saturation for holes. The predictions for the effective trapping probabilities for DOFZ material and for holes are probably overestimated.

If $Si_{FFCD}$ is a defect stable in time as the model supposes, in conditions of continuous irradiation, the accumulation of this defect will produce a strong degradation of the parameters of devices.

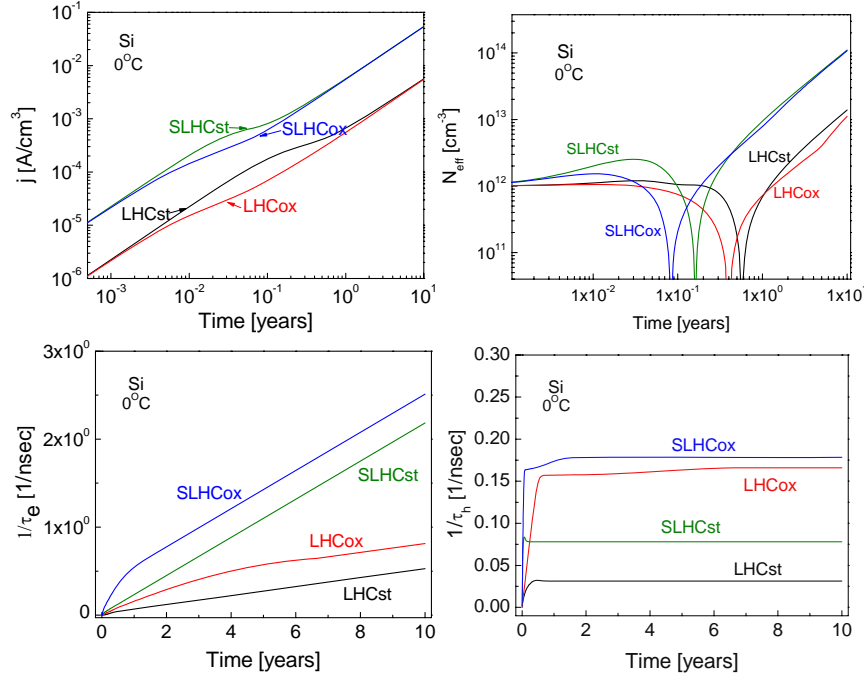

Fig. 5.
Time dependence of: a) volume density of the leakage current; b) Neff; and effective trapping probabilities for c) electrons and d) holes in the environment predicted for LHC and SLHC for standard and oxygenated silicon.

## 4. Summary

In this contribution, the correlations between microscopic mechanisms of degradation and macroscopic effects induced by irradiation, and the role of $Si_{FFCD}$ in these phenomena have been investigated. Because the background radiation field at LHC and SLHC will be an important source of production of defects, the expected degradation of silicon detectors has also been estimated.